\documentclass[12pt]{article}

\newcommand{\beg}{\begin{equation}}
\newcommand{\ene}{\end{equation}}

\begin{document}
\title{
\textsc{\bf Photon Antibunching,  Sub-Poisson Statistics and\\
Cauchy-Bunyakovsky and Bell's Inequalities}}

\author{
Igor V. Volovich
\\ $~~~$\\
\textsf{Steklov Mathematical Institute}\\
\textsf{Russian Academy of Sciences}\\
\textsf{Gubkin St. 8, 119991,  Moscow, Russia}\\
}
\date {~}
\maketitle
\begin{abstract}
We discuss some mathematical aspects of photon antibunching and sub-Poisson photon statistics.  It is known that Bell's inequalities for entangled states can be reduced to the Cauchy-Bunyakovsky inequalities. In this note some rigorous results on impossibility of classical hidden variables representations of certain quantum correlation functions are proved which are also based on the Cauchy-Bunyakovsky inequalities.

The difference $K$ between the variance and the mean as a measure of non-classicality of a state is discussed. For the classical case $K$ is nonnegative while for the $n$-particle state it is negative and moreover it  equals $-n$. The non-classicality of quantum states discussed here for the sub-Poisson statistics is different from another non-classicality  called entanglement
though both can be traced to the violation of the Cauchy-Bunyakovsky inequality.
\end{abstract}

\section{Introduction}
Photon antibunching is characteristic of a light field with photons more equally spaced than a coherent laser field,
e.g. light emitted from a single atom. It can also refer to sub-Poisson photon statistics, that is a photon number distribution for which the variance is less than the mean.

Photon antibunching and sub-Poisson photon statistics  reveal the quantum
nature of light and have been studied in many works on quantum optics \cite{MW}.

Here we consider some mathematical aspects of  the theory of such states. We give a rigorous formulation of sub-Poisson statistics and antibunching and prove the impossibility of
classical probabilistic representations of quantum correlation functions in these cases. It shows that the classical underlying random fields do not exist in the cases of sub-Poisson statistics and antibunching.

Note that impossibility of the existence of classical random
variable representation for some correlation functions of entangled states was rigorously proved by Bell,
see \cite{OV} for a recent discussion including the spatial dependence of entangled states.

It is known that Bell's inequalities for entangled states can be reduced to the Cauchy-Bunyakovsky inequalities \cite{Kit}. In this note some rigorous results on impossibility of classical hidden variables representations of certain quantum correlation functions are proved which are also based on the Cauchy-Bunyakovsky inequalities.

The difference $K$ between the variance and the mean as a measure of non-classicality of a state is discussed. For the classical case $K$ is nonnegative while for the $n$-particle state it is negative and moreover it  equals $-n$. The non-classicality of quantum states discussed here for the sub-Poisson statistics is different from another non-classicality  called entanglement
though both can be traced to the violation of the Cauchy-Bunyakovsky inequality.
\section{Sub-Poisson photon statistics}
 The sub-Poisson photon statistics is such  a photon number distribution for which the variance is less than the mean.
 Remind that for the Poisson statistics (for coherent states
 of light) they are equal. We consider a single-mode radiation
field.

 The variance of the photon number distribution is
\begin{equation}\label{var}
\langle \Delta n^2\rangle=\langle  n^2\rangle
-\langle  n\rangle^2,
\end{equation}
where
\begin{equation}\label{ndef}
n=a^*a,
\end{equation}
and the commutation relations for the
annihilation and creation operators $a$ and $a^*$ in
$L^2(R)$ are
\begin{equation}\label{cr}
[a,a^*]=1.
\end{equation}
We denote here $\langle  n\rangle=\langle\psi|  n|\psi\rangle$ where the vector $\psi$ is a unite vector in a dense domain of subspace of unite vectors in
$L^2(R)$.

By using the commutation relations one gets
\begin{equation}\label{var-2}
K\equiv\langle \Delta n^2\rangle-\langle n\rangle=\langle  a^{*2}a^2\rangle
-\langle  a^*a\rangle^2.
\end{equation}

Note that one has
\begin{equation}\label{var-2-1}
K=\langle  a^{*2}a^2\rangle
-\langle  a^*a\rangle^2=||a^2\psi||^2-||a\psi||^4.
\end{equation}
{\bf Definition.}  If for a unite vector $\psi$ one has
\begin{equation}\label{var-2-sP}
K=||a^2\psi||^2-||a\psi||^4<0,
\end{equation}
then the vector $\psi$ is called having the sub-Poisson statistics.

\subsection{$n$-particle states}

If $\psi_n$ is an $n$-particle state, $a\psi_n=\sqrt{n}\psi_{n-1}$, then one has
\begin{equation}\label{var-2-sP-p}
K=||a^2\psi_n||^2-||a\psi_n||^4=n(n-1)-n^2=-n<0, \,\,n=1,2,...
\end{equation}
In the case
\begin{equation}\label{var-2-sP-p-3}
\Phi=\sum_nc_n\psi_n,
\end{equation}
one gets
\begin{equation}\label{var-2-sP-p-u}
K=||a^2\Phi||^2-||a\Phi||^4=\sum_n n(n-1)x_n-(\sum_n nx_n)^2
\end{equation}
where
\begin{equation}\label{var-2-sP-p-j}
x_n=|c_n|^2,\,\,\sum_n x_n=1,\,\,x_n\geq 0.
\end{equation}
In particular for $\Phi=c_1\psi_1+c_2\psi_2$ one has $min K=-2$.

\subsection{Underlying classical random variables}
{\bf Definition.} We say that the annihilation and creation operators $a$ and $a^*$ admit an underlying classical random variable $\alpha,$ where $\alpha:\Omega\to C$, and $(\Omega, \sum,\mu)$ is a probability space,
if the following equality holds
\begin{equation}\label{class}
K=\langle\psi|  a^{*2}a^2|\psi\rangle-\langle\psi|  a^*a|\psi\rangle^2=E(\alpha^{*2}\alpha^2)-(E(\alpha^*\alpha))^2,
\end{equation}
for any unite vector $\psi$  in a dense domain of subspace of unite vectors in
$L^2(R)$. Here $E(\alpha^{*2}\alpha^2)=\int_{\Omega}|\alpha|^{4}d\mu$
and
$E(\alpha^*\alpha)=\int_{\Omega}|\alpha|^2 d\mu$. The probabilistic measure $\mu$ may depend from $\psi$.

{\bf Remark.} We could require
\begin{equation}\label{class-1}
\langle\psi|  a^{*m}a^l|\psi\rangle=E(\alpha^{*m}\alpha^l),\,\,m,l=0,1,2,...
\end{equation}

If there exists an underlying random variable
then the variance would have to be greater than or equal to the mean. Indeed, then the right hand side of (\ref{var-2}) would be equal to
\begin{equation}\label{class-2}
E(\alpha^{*2}\alpha^2)-(E(\alpha^*\alpha))^2\geq 0.
\end{equation}
It is nonnegative, by the Cauchy-Bunyakovsky  inequality since
$\int fd\mu \leq (\int f^2d\mu)^{1/2}$, where $f=|\alpha|^2$.

We have the following

{\bf Theorem.} {\it The annihilation and creation operators $a$ and $a^*$ do not admit an underlying classical random variable. }

{\bf Proof.} We have to prove that the representation (\ref{class}) can not be valid for any $\psi$ from  a dense domain in
$L^2(R)$. We already have shown that for the $n$-particle states $K$ is negative, $K=-n$, see (\ref{var-2-sP-p}).
This contradicts to (\ref{class-2}) if there is
an underlying random variable.
 Theorem is proved.

\section{ Antibunching}

Antibunching is the violation of the inequality
\begin{equation}\label{in}
P(\tau)\leq P(0), \,\, \tau\geq 0.
\end{equation}
Here
\begin{equation}
P(\tau)=\langle\psi|E^-({\bf r},t)E^-({\bf r},t+\tau)E^+({\bf r},t+\tau)E^+({\bf r},t)|\psi\rangle,
\end{equation}
we consider stationary states $\psi$ and do not indicate
dependence from the space position of the detector
${\bf r}$. For a consideration of the dependence from the spatial variable ${\bf r}$, see \cite{OV}. 

The inequality (\ref{in}) can be understood in the following way. If $I(t)$ is a (classical) stationary random process
(in particular, $E I(t)=E I(0)$) then, due to the Schwarz inequality, one has:
\begin{equation}\label{Sw}
E(I(t)I(t+\tau))\,\leq\, (E I^2(t))^{1/2}(EI^2(t+\tau))^{1/2}=EI^2(0).
\end{equation}
The inequality (\ref{Sw}) is the same as (\ref{in}) if we set
\begin{equation}\label{why}
P(\tau)=E(I(t)I(t+\tau)).
\end{equation}
We have proved the following

{\bf Theorem.} {\it If for some state $\psi$ and for some $\tau >0$ one has
\begin{equation}\label{anti}
P(\tau)> P(0),
\end{equation}
then this state does not admit a classical description.}

In this situation one has antibunching. Such states $\psi$ are called nonclassical states.

\section{Conclusions}
In this note some non-classical properties of quantum states related with the violation of the Cauchy-Bunyakovsky inequality have been considered. Since this inequality is the key ingredient also in the proof of the Bell inequalities \cite{Kit} one can say that the violation of the  Cauchy-Bunyakovsky inequality is responsible for non-classical properties of quantum states such as entanglement, antibunching and sub-Poisson statistics.
The spacetime dependence of entangled states is considered
in \cite{OV}. It would be interesting to study the spacetime dependence of non-classical states discussed in this note.

Photon antibunching was observed in experiments with resonance fluorescence,  \cite{MW}.
It would be interesting to study the measure of non-classicality $K$ as a function from $x_n$, see (\ref{var-2-sP-p-u}), and also to consider a  generalization to the density operators.

\section{Acknowledgments}
I am grateful to
I. Arefeva,  A. Khrennikov and B. Nilsson
for useful comments. The work is partially supported by grants  RFFI 11-01-00828-a and NS 7675.2010.1 .

\end{document}